\documentclass[prd,twocolumn,showpacs,preprintnumbers,tightenlines,amssymb,
               eqsecnum,superscriptaddress,floatfix]{revtex4}
\usepackage{epsfig,hyperref,dcolumn}
\newcommand{\Real}{\mathbb{R}}

\begin{document}

\title{No naked singularities in homogeneous, spherically symmetric bubble spacetimes?}

\author{Olivier Sarbach and Luis Lehner}
\affiliation{Department of Physics and Astronomy, Louisiana State
University, Baton Rouge, LA 70810, U.S.A.} 

\begin{abstract}
We study the evolution of bubble spacetimes in vacuum and
electrovac scenarios by numerical means. We find strong
 evidence against the formation of
naked singularities in (i) scenarios with negative masses displaying
initially collapsing conditions and (ii) scenarios with negative
masses displaying initially expanding conditions, previously reported
to give rise to such singularities. Additionally, we show that the
presence of strong gauge fields implies that an initially collapsing bubble
bounces back and expands. By fine-tuning the strength of the gauge field
we find that the solution approaches a static bubble solution.
\end{abstract}


\maketitle

\section{Introduction}

Within five-dimensional Kaluza-Klein theory, negative energy initial
data configuration at a moment of time symmetry have been presented in
the past \cite{brillpfister,brillhorowitz}. Among the reasons for
considering negative energy solutions is that naked singularities are
associated to them. Therefore studying such solutions is an attractive
problem to further test the cosmic censorship
conjecture~\cite{penrose}.  Additionally, as these bubbles can be
obtained via semiclassical tunneling (following Witten's
construction~\cite{witten,aharony}), it is important to understand
their dynamical behavior. To obtain a better feeling for the dynamics
of the solutions presented in \cite{brillhorowitz}, which describe a
bubble spacetime, Corley and Jacobson \cite{corleyjacobson} analyzed
the initial acceleration of the bubble (which can be calculated
analytically with the information contained on a single hypersurface.)
They found that negative mass bubbles start out expanding and
consequently argued a naked singularity would be unlikely. This
conjecture is backed by the calculations at the initial hypersurface
together with the intuition gained from a time symmetric
scenario. Namely, that if the bubble were to reverse its behavior it
will go through another time-symmetric phase\footnote{However, we will
see that one can encounter situations in which a bubble reverses its
behavior without passing through a moment of time symmetry.}, which
would also suggest the bubble should expand (assuming the solution at
this time of symmetry still belongs to the family presented in
\cite{brillhorowitz}).

Recently however, a numerical study was presented~\cite{shinkai}
which computed explicitly the solution to the future of the initial,
time-symmetric, initial data. In the negative mass case, the solution
found appeared to indicate a very puzzling development. Namely, that
as the bubble expanded, it would encounter a naked singularity on its
way.

In the present work we reexamine the dynamical behavior of the data
presented in \cite{brillhorowitz} via numerical simulations, and
further study a more general case where the initial data describes an
initially collapsing bubble with negative mass. We restrict our simulations
to the context of homogeneous spherically symmetric spacetimes.
To summarize our findings, we see that
negative mass bubbles will not lead to a naked singularity. By
choosing different members of the initial data family, these could
start out either expanding or collapsing. In the former case, the
bubble continues on expanding while in the latter case, the bubble
bounces back without ever collapsing. We additionally study the case
of a positive mass bubble which starts out collapsing. Here, depending
on the strength of the gauge field at the initial hypersurface, the
bubble collapses to a black string or bounces back to expand
``forever'' (i.e. for as long as we let the simulation run).
Furthermore, these two distinct possibilities give rise to critical
behavior on the threshold of collapse or expansion.

\section{Set up: Bubble data}

We consider a generalization of the time symmetric family of initial
data presented in \cite{brillhorowitz}. We start with a spacetime
endowed with the metric
\begin{equation}
ds^2 = -dt^2 + U(r) dz^2 + \frac{dr^2}{U(r)} + r^2 d\Omega^2 ,
\label{Eq:BubbleMetric}
\end{equation}
where $d\Omega^2 = d\vartheta^2 + \sin^2\vartheta d\varphi^2$ is the
standard metric on the unit two-sphere $S^2$ and $U(r)$ is a smooth
function that has a regular root at some $r=r_+ > 0$, is everywhere
positive for $r > r_+$ and converges to one as $r\rightarrow\infty$.
The coordinate $z$ parametrizes the extra dimension $S^1$ which has
the period $P = 4\pi/U'(r_+)$. The resulting spacetime $\{ t,z,r\geq
r_+,\vartheta,\varphi \}$ constitutes a regular manifold with the
topology $\Real\times\Real^2\times S^2$.  The bubble is located where
the circumference of the extra dimension shrinks to zero, that is, at
$r=r_+$.

Additionally, we consider the presence of a {\bf U(1)} gauge field in the
following form,
\begin{equation}
A_a dx^a = k (r_+^{-n} - r^{-n}) dz,
\end {equation} 
with $k$ an arbitrary constant and $n$ an integer greater than
one.  It is important to point out two
things. First, the gauge field here considered has a more generic
dependence than those previously considered in \cite{brillhorowitz},
where only the case $n=2$ was discussed. Second, the symmetries of the
problem would a priori allow for the gauge field to have non-trivial
$t$ and $r$ components.  However, it follows from Maxwell's equations
that those contributions would lead to a Coulomb-like electric field
which is divergent at the location of the bubble. For this reason we
set those components of $A_a$ to zero.

In the time-symmetric case, initial data satisfying the Hamiltonian
constraint obeys
\begin{equation}
U(r) = 1 - \frac{m}{r} + \frac{b}{r^2} - \frac{\tilde k^2}{r^{2n}}\, ,
\end{equation} 
where $\tilde{k} \equiv k n/\sqrt{(n-1)(2n-1)}$. The parameter $m$ is
related to the ADM mass via $M_{ADM} = m/4$.

The fact that the bubble be located at $r=r_+$ requires that $0 =
U(r_+) = 1 - \bar{m} + \bar{b} - \bar{k}^2$, where $\bar{m} \equiv
m/r_+$, $\bar{b} \equiv b/r_+^2$, $\bar{k} \equiv \tilde{k}/r_+^n\,$. We
also require
\begin{equation}
0 < r_+ U'(r_+) = 2 - \bar{m} + 2(n-1)\bar{k}^2
\label{Eq:RegHor}
\end{equation}
and avoid the conical singularity at $r=r_+$ by fixing the period of
$z$ to $P = 4\pi/U'(r_+)$. Repeating the analysis in
\cite{corleyjacobson} one finds that the initial acceleration of the
bubble area is given by
\begin{equation}
\ddot{A} = 8\pi\left[ 1 - \bar{m} - \frac{4\bar{k}^2}{3}(n-1)(n-2) \right].
\label{Eq:ddotA}
\end{equation}
For $n=2$ this agrees with the formula found by Corley and Jacobson,
and as they discussed, for negative mass bubbles, the initial
acceleration is positive, and only if chosen appropriately can a
positive mass bubble have negative initial acceleration. As we will
see later, in the vacuum case, this acceleration remains negative
leading to a collapse of the bubble and the formation of a black
string. In the non-vacuum case however, the strength of the gauge
field can modify this behavior completely. For weak enough $k$ the
bubble continues to collapse whereas when $k$ is large the bubble area
bounces back and expands. Furthermore, by fine-tuning the parameter
$k$ we will show that the solution approaches a static bubble solution
to the five-dimensional Einstein-Maxwell equations.

We also consider non-vacuum initial data with $n > 2$ which allow for
richer scenarios. A particularly interesting case is that describing
an initially negative acceleration with negative mass. If this were
possible, and the bubble continued to collapse, this would likely give
rise to a naked singularity. Thus, we are interested in data for which
$\ddot{A} < 0$. In view of the inequality (\ref{Eq:RegHor}) and
Eq. (\ref{Eq:ddotA}) this means that
\begin{equation}
2 + 2(n-1)\bar{k}^2 > \bar{m} > 1 - \frac{4\bar{k}^2}{3}(n-1)(n-2).
\end{equation}
From this we see that for $n = 2$ the mass can only be positive. In
contrast to this, for $n > 2$ and $\bar{k}$ large enough such that
$4\bar{k}^2(n-1)(n-2) > 3$, negative mass bubbles can be considered.

\section{Numerical implementation}

A detailed description of the numerical techniques used in our code
will be presented elsewhere \cite{threetenors}. Here we point out its
salient features.
\begin{enumerate}
\item 
The equations are written in first oder symmetric hyperbolic form,
with a gauge choice that is such that the characteristic speeds are
given by $0$, $1$ or $-1$. These properties facilitate achieving a
stable implementation of the equations.
\item
Regularity conditions are enforced at the bubble and employed to
design a numerical scheme that satisfies a discrete energy estimate
\cite{lsu1,lsu2}. This discrete energy estimate guarantees numerical
stability for a related toy model problem.
\item
Constraint preserving boundary conditions are enforced at the outer
boundary to ensure no constraint violating modes are fed through
it~\cite{calabreselehnertiglio,sarbachetal}.
\item 
The equations are implemented using second order accurate finite
difference techniques. A non-uniform radial coordinate is employed to
improve the resolution near the bubble. For all cases presented here,
the radial grid covered $r \in [1,20]$ with a total of $8000$
points. The outer boundary's location is chosen so as to have {\it
always} less than a crossing time in our evolutions, guaranteeing
outer boundary conditions (though consistent with the constraints)
will not affect quantities at the bubble. Second order convergence is
checked in all runs. As an example, the convergence of some of the
simulations representing expanding bubbles is illustrated in
Fig. \ref{fig:conv1}. Similar behavior is observed in all
cases\footnote{As mentioned, in the cases with negative mass, our
results disagree with those presented in \cite{shinkai}. We
believe that the reason for the disagreement is that the numerical
simulations presented in \cite{shinkai} were contaminated by ill-posed
modes present in the weakly hyperbolic form of the equations and the
implementation of the boundary conditions used there. The
understanding of these issues and how to deal with them has advanced
considerably in the past few years. In particular, points 1-3 mentioned
in section III 
are major differences between the two implementations.}.
\end{enumerate}

\section{Results}

\subsection{Brill-Horowitz initially expanding case: $(n=2)$}

\begin{figure}[h]
\centerline{\epsfxsize=200pt\epsfysize=130pt\epsfbox{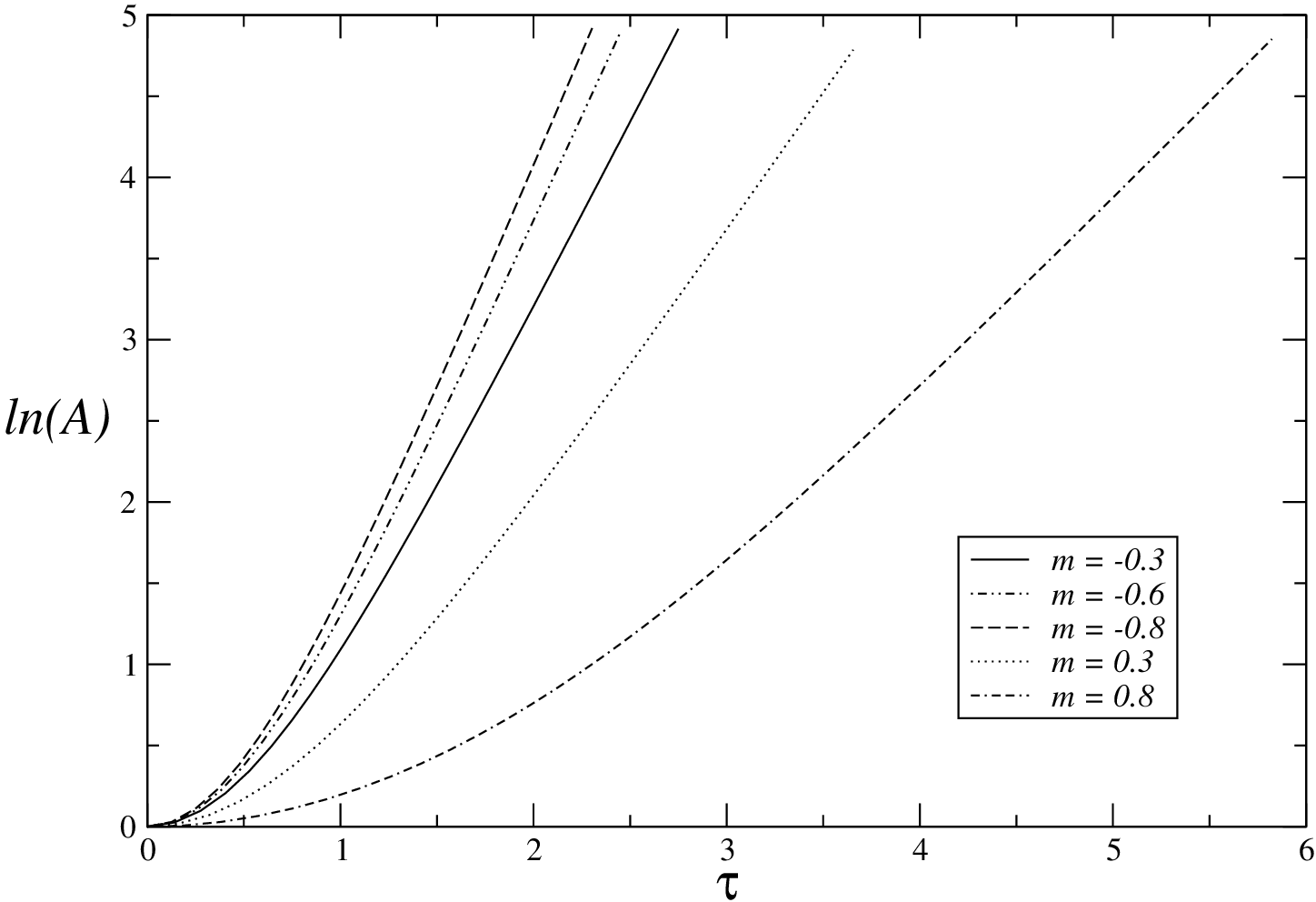}}
\caption{Bubble area vs. proper time at the bubble. In this and the
following plots, we set $r_+ = 1$ (by choosing $b$ appropriately). The figure shows four illustrative
examples of bubbles whose initial acceleration is positive. As it is
evident, the expansion of the bubble continues and the difference is
the rate of the exponential expansion. The relative error in these
curves is estimated to be well below 0.001\%.  \\ \\ }
\label{fig:area}
\end{figure}

\begin{figure}[h]
\centerline{\epsfxsize=200pt\epsfysize=150pt\epsfbox{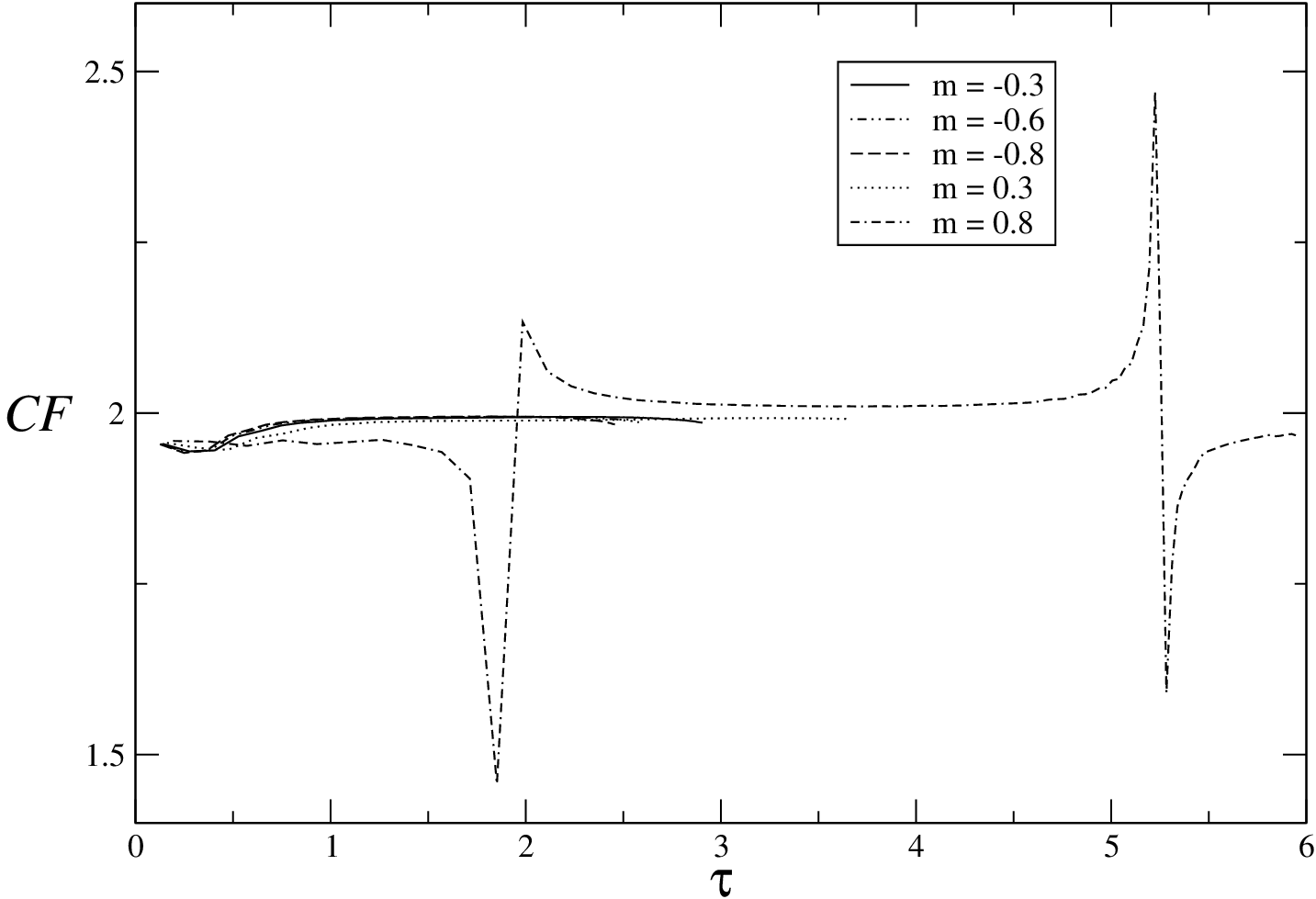}}
\caption{Convergence factor (CF) vs. proper time for the cases illustrated
in Fig. \ref{fig:area}. This factor, defined as $CF=log_2(
||A(4\Delta)-A(2\Delta)||/||A(2\Delta)-A(\Delta)|| )$ should give the
value of $2$ for a second order accurate implementation.  Through
these runs the grid spacing is defined as $\Delta = 3.75 \times 10^{-3}$
Clearly, the plot shows that second order convergence has been
obtained throughout the runs (the peaks are instances with $0/0$ where
the expression for $CF$ is ill-defined).}
\label{fig:conv1}
\end{figure}

Here we evolve the Brill-Horowitz initial data in the case of
vanishing gauge field. The bubble area $A$ as a function of the proper
time $\tau$ at the bubble is shown in Fig. \ref{fig:area} for
different values of the mass parameter $m$. (Figure \ref{fig:conv1}
illustrates the convergence of the solution throughout the
evolution). As expected, the lower the mass of the initial
configuration, the faster the expansion.  Empirically, and for the
parameter ranges used in our runs, we found that at late times the
expansion rate is given by
\begin{equation}
\frac{\dot{A}}{A} \approx \frac{2-\bar{m}}{r_+(\tau=0)}\, ,
\label{simplerelation}
\end{equation}
where a dot denotes the derivative with respect to proper time
$\tau$. In particular this approximation is valid for the bubble solution
exhibited by Witten \cite{witten} which describes the time evolution
in the case $\bar{m}=0$. We monitored several curvature invariant
quantities (the Kretschmann invariant in particular) for our numerical
spacetimes and found no sign of divergence.

\subsection{Brill-Horowitz initially collapsing case: $(n=2)$}

Next, we analyze the Brill-Horowitz initial data for the case in which
the bubble is initially collapsing (notice that for $n=2$ this implies
that the ADM mass is positive). While our numerical simulations reveal
that in the absence of the gauge field such a bubble continues to
collapse, we also show that when the gauge field is strong enough, the
bubble shrinks at a rate which decreases with time and then bounces
back.

\begin{figure}[h]
\centerline{\epsfxsize=200pt\epsfysize=130pt\epsfbox{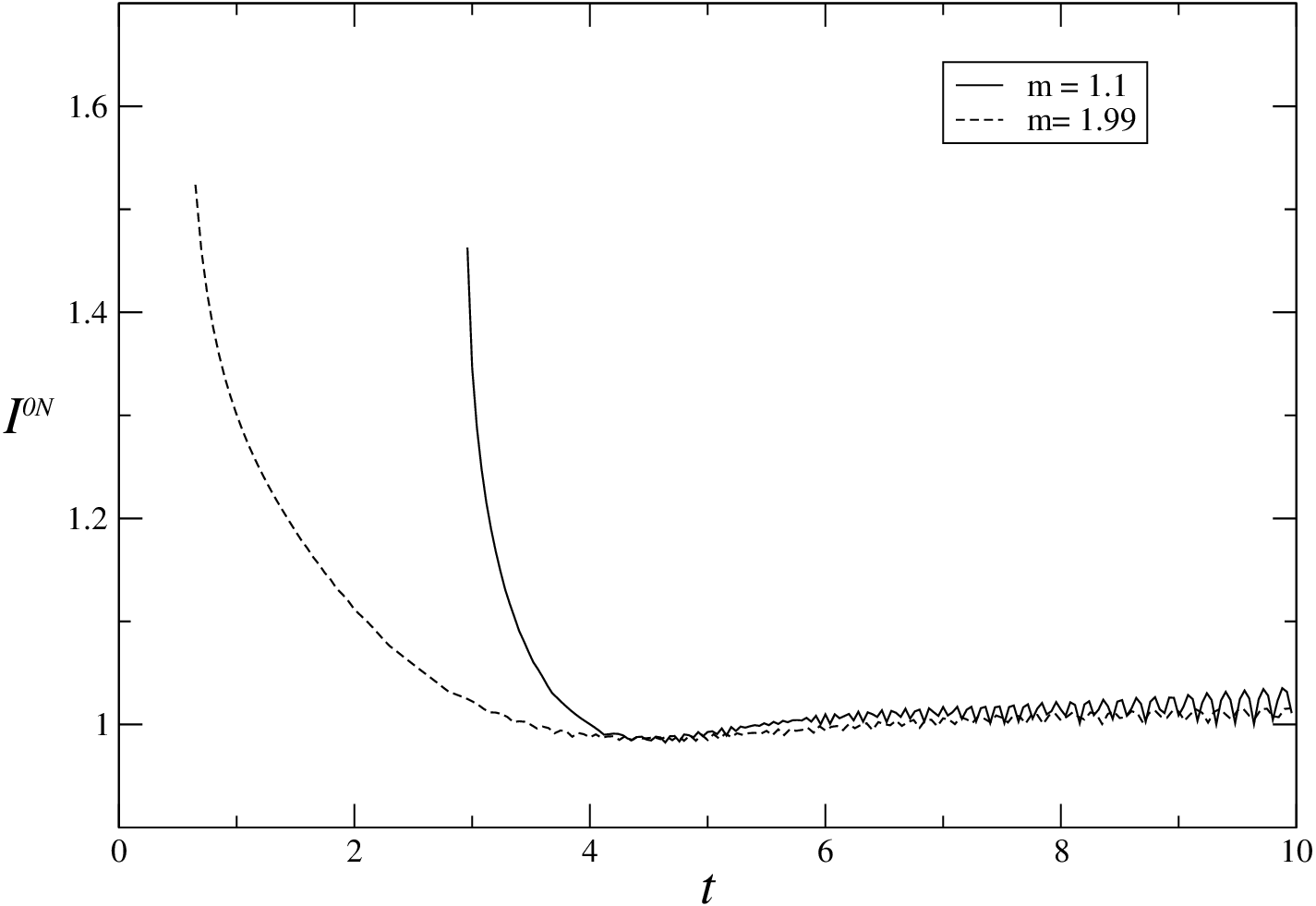}}
\caption{Rescaled Kretschmann invariant $I R_{AH}^4/12$ 
vs. asymptotic time for $m=1.1$ (solid line) and $1.99$ (dashed line). The first non-zero
values of the lines mark the formation of the apparent horizon. At late
times, both lines approach the value of $1$ suggesting a black
string has formed.}
\label{fig:blackstring}
\end{figure}

Obviously, if the collapse trend were not halted, a singularity should
form at the origin. Since the ADM mass is positive, one expects this
singularity to be hidden behind an event horizon, and one should
obtain a black string. In fact, for the solutions which are initially
collapsing and which have vanishing gauge field, we observe the
formation of an apparent horizon. Furthermore, we compute the
curvature invariant quantity $I R_{AH}^4$ at the apparent horizon (as
discussed in \cite{choptuikblackstring}), where $I = R_{abcd}R^{abcd}$
is the Kretschmann invariant and $R_{AH}$ the areal radius of the
horizon. For a neutral black string, this invariant is $12$.  Figure
\ref{fig:blackstring} shows how this value is attained after the
apparent horizon forms for representative vacuum cases (with $m=1.1$
and $m=1.99$), thus providing strong evidence for the formation of a
black string. (We also find apparent horizons forming, or already
present in the initial data, when a nontrivial gauge field is
considered and the parameters are appropriately chosen. The
corresponding bubbles will likely collapse to charged black strings
and will be analyzed elsewhere~\cite{threetenors}.)\\ \\

\begin{figure}[h]
\centerline{\epsfxsize=200pt\epsfysize=120pt\epsfbox{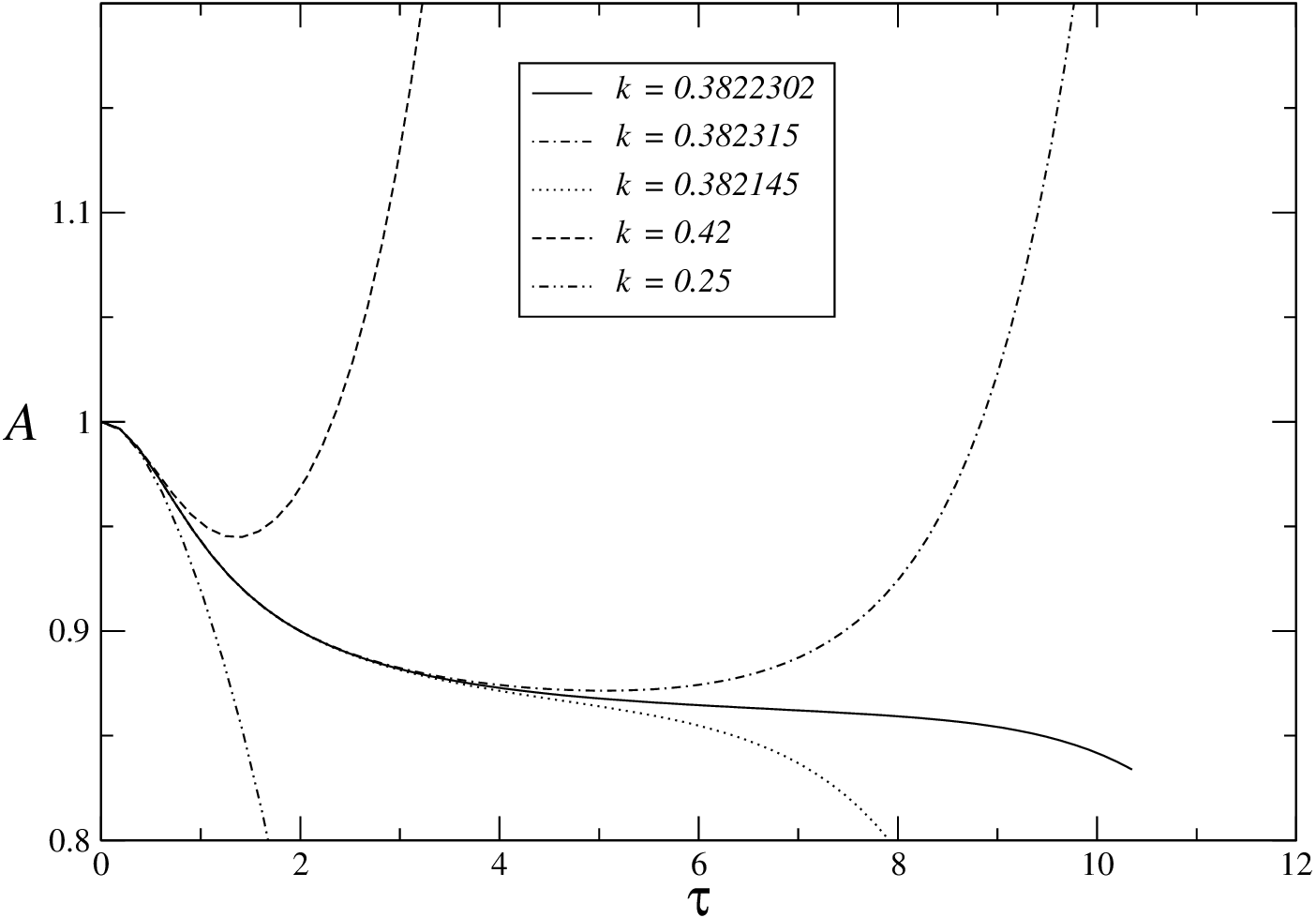}}
\caption{Area values vs. proper time at the bubble for different
values of $k$ and $m=1.1$. By tuning the value of $k$ appropriately,
the amount of time that the area remains fairly constant can be
extended for as long as desired.}
\label{fig:criticalonset}
\end{figure}

As mentioned, for strong enough gauge fields, the previously described
dynamics is severely affected. Figure \ref{fig:criticalonset} shows
the bubble area vs. proper time for different values of $k$. For large
values of $k$ the bubble ``bounces'' back and expands while for
small ones the bubble collapses. There is a natural transition region
separating these two possibilities. Tuning the value of $k$ one can
reveal an associated critical phenomena, the `critical solution' being
a member of the family of static solutions given by
\begin{eqnarray}
&& ds^2 = -V(r) dt^2 + \frac{V(r)}{U(r)} dr^2 + \frac{U(r)}{V(r)^2} dz^2 
 + r^2 V(r) d\Omega^2, \nonumber\\
&& A_a dx^a = \pm\frac{1}{2}\sqrt{3\left(\frac{r_+}{r_-} - 1 \right)}\,\frac{dz}{V(r)}\, ,
\nonumber
\end{eqnarray}
where $V(r) = 1 - r_-/r$ and $U(r) = 1 - r_+/r$. These solutions can
be obtained by a double analytic expansion of the black hole solutions
described by Eqs. (2.13) and (2.15) of
Ref.~\cite{horowitzmaeda_solution} followed by the transformations
$r_- \mapsto -r_-$, $r \mapsto r - r_-$, $r_+ \mapsto r_+ - r_-$. The
parameters $r_-$ and $r_+$ $(> r_-)$ are related to the period of the
$z$ coordinate and to the ADM mass via $P = 4\pi r_+(1 -
r_-/r_+)^{3/2}$ and $M_{ADM} = r_+/4$. Since these parameters are
conserved throughout evolution, the initial data fixes the member of
the static family the dynamical solution approaches to. Details of
the critical behavior exhibited by this system will be presented
in~\cite{threetenors}.\\ \\

\subsection{Collapsing negative mass case}
\begin{figure}[h]
\centerline{\epsfxsize=240pt\epsfysize=150pt\epsfbox{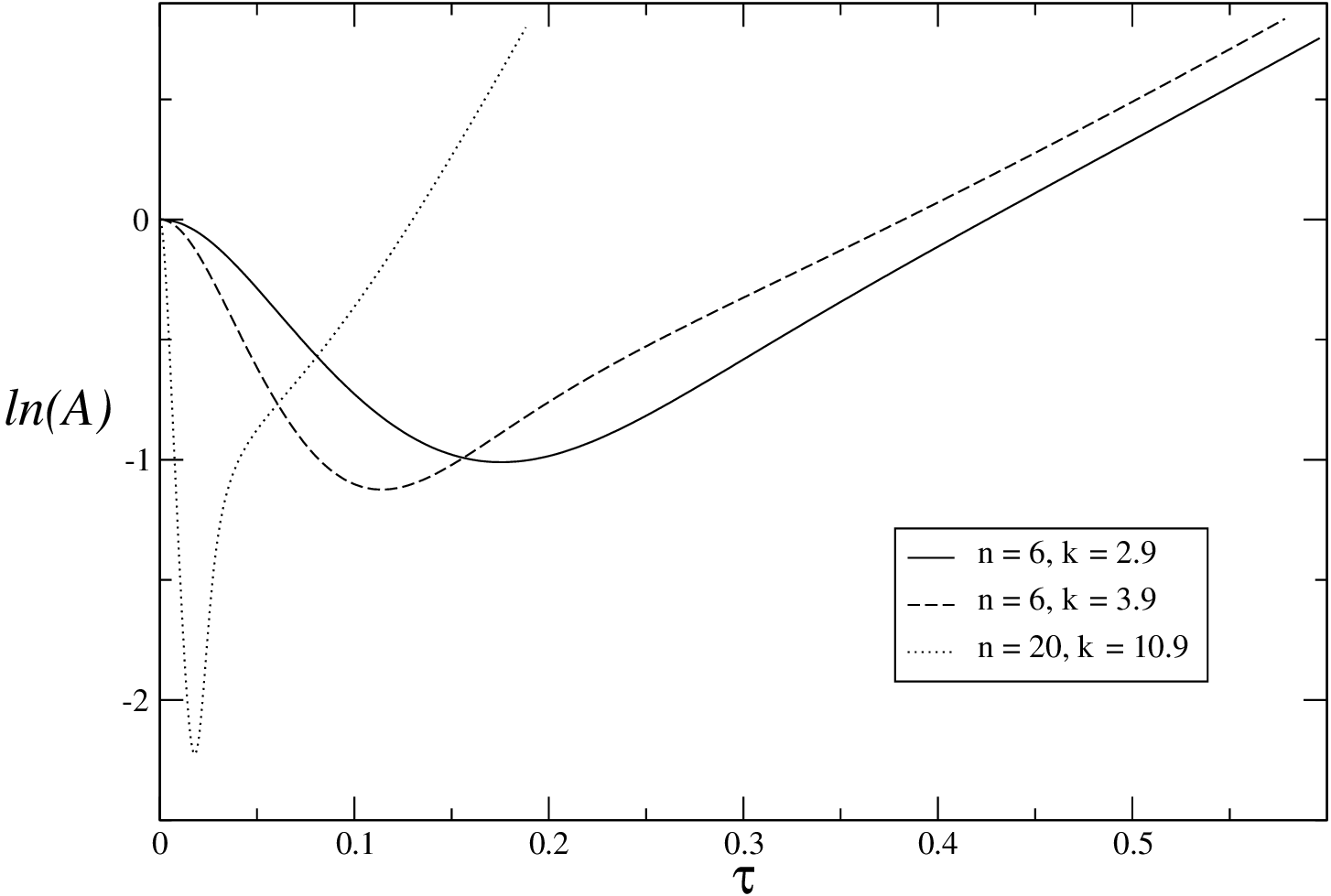}}
\caption{Bubble area vs. proper time at the bubble. The figure shows
three illustrative examples of bubble with negative mass ($m = -0.1$
each) whose initial acceleration is negative. As it is evident, the
collapse of the bubble is halted and the trend is completely
reversed. The error in these curves is estimated to be well below
0.001\%.\\
\\}
\label{fig:notnakedIII}
\end{figure}

We here restrict to cases with negative masses that start out
collapsing. Interestingly enough we find that even when starting with
large initial negative accelerations, which in turn make the bubble
shrink in size to very tiny values, it bounces back without ever
collapsing into a naked singularity. As an example, Fig.
\ref{fig:notnakedIII} shows the bubble's area versus time for
different values of $n$ and $k$. The initially collapsing bubbles
decrease in size in a noticeable way but this trend is halted and the
bubbles bounce back and expand. Although we have not found a simple
law as that in Eq. (\ref{simplerelation}), clearly the bubbles expand
exponentially fast. Therefore, it seems not to be possible to
``destroy'' the bubble and create a naked singularity. This situation
is somewhat similar to the scenarios where one tries to ``destroy'' an
extremal Reissner-Nordstr\"om black hole by attempting to drop into it
a test particle with high charge to mass ratio. There, the
electrostatic repulsion prevents the particle from entering the hole
\cite{wald}.

\section{Conclusions}
We have performed numerical simulations to investigate the dynamics of
spherically symmetric, homogeneous bubble spacetimes. These have
revealed several interesting features.

In the vacuum case, negative mass bubbles do not form singularities,
rather they continue on expanding and the asymptotic rate of expansion
obeys the simple relation (\ref{simplerelation}). Positive mass cases,
as pointed out previously can start out either expanding or
contracting. In the former case, they continue on expanding and its
asymptotic rate obeys the same simple relation.  For the collapsing
case we have evidences that the bubble will collapse to a black
string. It is interesting to note here that the ratio of a circle at
infinity over the ADM Mass is given by
\begin{equation}
\frac{P}{M_{ADM}} = \frac{16 \pi}{\bar m (2-\bar m)} \, . 
\end{equation}
Since $0 < \bar m < 2$ this value is always larger than that needed
for the Gregory-Laflamme ``instability''\cite{gregorylaflamme}.
Thus, a slight $z$-dependent perturbation of the scenario considered
here could indeed give rise to a naked singularity. This would require
infinite time to appear as shown in \cite{horowitzblackstring}, but
certainly would give rise to rich dynamics as illustrated in
\cite{choptuikblackstring}.

In the non-vacuum case, the presence of a gauge field gives rise to
additional features. First, for the positive mass case, it can
``halt'' the collapse and subsequently induce an expansion.
Furthermore, by fine-tuning the strength of the field, the solution
approaches a static bubble. This critical solution separates the
collapsing bubbles (which, likely, collapse to a charged black string)
from the one that expand at late times. In a subsequent
paper~\cite{threetenors} we will show that the dynamical behavior near
the static bubble exhibits critical phenomena. Note that clearly the
solution does encounter another moment of time symmetry at the
critical value. However, for field strengths larger than the critical
one, this is not the case. The bubble bounces but the solution does
not go through a time-symmetric phase.

Last, the gauge field can also make {\it negative} mass bubbles start
out collapsing if its radial dependence is steeper than that originally
considered in \cite{brillhorowitz}. Intuitively this happens due to
the contribution of the energy in the gauge field overcoming the
repulsive gravitational effect of the negative mass bubble. However,
as part of the field is radiated away, the collapse is halted and a
subsequent expansion results.  By choosing both the strength of the
field or its radial dependence appropriately (with $n>2$), we could
make the bubble start out collapsing at a very rapid rate. In these
cases, the bubble shrinks to very small sizes but eventually bounces
back without forming a naked singularity.

\begin{acknowledgments}
We thank Robert Myers for calling this problem to our attention and
for pointing out the influence a steeper radial dependence in the
gauge field could have. Additionally, we thank him and Gary Horowitz
for many interesting discussions. We acknowlege the LSU gravity
group for comments on the manuscript. We benefited from the hospitality of
the Kavli Institute for Theoretical Physics. L.L. thanks the Perimeter
Institute for Theoretical Physics for hospitality.  This research was
supported in part by the NSF under grant numbers NSF-PHY0244335 to Louisiana
State University, PHY99-07949 to the University of California at Santa Barbara
and by the Horace Hearne Institute for Theoretical
Physics. L.L. has been partially supported by the Alfred P. Sloan Foundation.
\end{acknowledgments}

\end{document}